\documentstyle[11pt,aaspp4]{article}

\begin{document}

\title{Effect of the Global Rotation of the Universe on \\the Formation
of Galaxies}

\author{Li-Xin Li\altaffilmark{1}}
\affil{Institute of Theoretical Physics and Department of Physics\\
Beijing Normal University, Beijing 100875, China}

\altaffiltext{1}{E-mail address: lixin@eden.rutgers.edu}

\begin{abstract}
The effect of the global rotation of the universe on the formation
of galaxies is investigated. It is found that the global rotation
provides a natural origin for the rotation of galaxies, and the
morphology of the objects formed from gravitational instability in a
rotating and expanding universe depends on the amplitude of the
density fluctuation, different values of the amplitude of the
fluctuation lead to the formation of elliptical galaxies, spiral
galaxies, and walls. The global rotation gives a natural explanation
of the empirical relation between the angular momentum and mass of
galaxies: $J\propto M^{5/3}$. The present angular velocity of the universe
is estimated, which is $\sim 10^{-13}$ rad yr$^{-1}$.
\end{abstract}



\section*{}
In a homogeneous universe which is more general than the Friedmann
model, the matter may not only expand but also rotate relative to
local gyroscopes. The rotation of the matter in the universe as a
whole is usually called the global rotation of the universe, which has
been investigated by many scientists (\cite{gam46,god49} \& 1990,
\cite{ell71,obu92,kor96}). People usually think that
the observed isotropy of the cosmic microwave background (CMB)
strongly restricts the possible value of the angular velocity of the
universe (\cite{col73,haw74}). However, when more 
general cosmological models are
considered, the restriction may be much looser (\cite{mat97}). For the
Bianchi IX models the more realistic limits are thought to be of the order
$10^{-12}$ rad yr$^{-1}$ (\cite{ciu95}). And more, some
recent investigations
reveal that there are a wide class of viable cosmological models for
which the global rotation does not influence the isotropy of CMB at
all and it is the shear which may affect the isotropy of CMB
(\cite{obu90,kor91} \& 1996, \cite{pav95}). 
Therefore it is significant to investigate the cosmic effects of
the global rotation further.

Birch (1982) has discovered the asymmetric distribution of the angles of the
rotation of polarization vector of 132 radio sources and has tried to
explain it via the global rotation. Korotky and Obukhov (1994) have
applied the global rotation to explain the observed periodic
distribution of galaxies on the large scale (\cite{broa90}). Though the
result may be controversial (\cite{phi83,bir83}), 
it is sufficient to show that the
global rotation may be relevant to some important observational
phenomena.

Here I try to investigate the effect of the global rotation
on the formation of galaxies. I show that in a rotating universe
galaxies automatically get the angular momentum when they form due to
the conservation of angular momentum, which gives a natural
interpretation of the rotation of galaxies. Such an idea has been
postulated previously by some people (\cite{gam46,god90,col73}). 
However, detail analysis
has not been made and it has been worried that this may lead to that
the orientation of galaxies should be aligned in some direction which
seems contrary with the observations. But the anisotropy 
in the distribution of the orientation of galaxies has been
found at different levels (\cite{mac85,djo87,sug95,hu95})
and a pronounced anisotropy has been
found recently (\cite{par94}) though the origin of the anisotropy is still in
argument (\cite{fli95}). (In the end of this paper I will show that due to the
irregularity of the shape of the proto-galaxies the distribution of the
orientation of galaxies may be somewhat random which makes it
difficult to measure the correlation of the orientation of
galaxies). In this paper I
derive a correlation between the angular momentum $J$ and mass $M$ of
galaxies, which is consistent with the empirical relation
(\cite{bro63,oze67,bur75,tri88})
\begin{eqnarray}
    J\propto M^{5/3}.
\end{eqnarray}
Such an empirical relation was usually explained via the virial
theorem with the assumption that galaxies have constant density
(\cite{oze67}),
but why such an assumption should hold was not explained. Here I show
that the global rotation may give a natural explanation of this
relation which does not require the assumption of constant
density. The present value of the angular velocity of the global
rotation is estimated from the statistical analysis of the correlation
between the angular momentum and mass of galaxies. The result is just
within the limits of CMB for the Bianchi IX models obtained by Matzner
and cited by Ciufolini and Wheeler (1995). The value of the angular
velocity obtained is of the same order as that obtained by Birch (1982)
and consistent with that obtained by Obukhov (1992), Korotky and Obukhov
(1994). The relation between the primordial density fluctuation and the
formation of galaxies in a rotating and expanding universe is also
discussed. it is found that the morphology of the objects formed
depends on the amplitude of the density fluctuation, different values
of the fluctuations lead to the formation of elliptical galaxies,
spiral galaxies, and walls.

The motion of the fluid in the universe can be described by the volume
expansion scalar $\Theta$, the rotation tensor $\omega_{ab}$, and the
shear tensor $\sigma_{ab}$. The homogeneous rotation of the fluid as a
whole is the global rotation of the universe. If the
fluid is the perfect fluid with the stress-energy tensor
$T_{ab}=(\rho+p)u_au_b+pg_{ab}$ ($\rho$ is the mass density and $p$ is
the pressure), with the Einstein equation the Raychaudhuri 
equation describing the relation among
$\Theta$, $\omega_{ab}$, and $\sigma_{ab}$ can be written as 
\begin{eqnarray}
    -\nabla_a A^a + \dot\Theta + {1\over 3}\Theta^2 + 2(\sigma^2-\omega^2)
    = -4\pi G(\rho+3p),
\end{eqnarray}
where $A^a=u^b\nabla_b u^a$ is the acceleration vector, the dot
denotes the derivative $u^a\nabla_a$, and
$\omega^2\equiv\omega_{ab}\omega^{ab}/2$,
$\sigma^2\equiv\sigma_{ab}\sigma^{ab}/2$ (\cite{ciu95}). $\omega$ is also called
the scalar angular velocity. The most important cases for perfect
fluid are dust and radiation for that the
universe is dominated by dust when $z<z_{eq}$ and dominated by
radiation when $z>z_{eq}$ where $z_{eq}\sim 10^4$ is the redshift
($z\equiv a_0/a-1$ where $a$ is the scale function defined by
$\Omega=3\dot a/a$ and the index ``0'' denotes the value at the
present epoch) when the mass densities of dust and radiation are
equal. It has been shown that the spatially homogeneous, rotating, and
expanding universes filled with perfect fluid must have a
non-vanishing shear (\cite{kin73,ray79}). 
However, it seems reasonable to assume
that $\sigma$ is sufficiently small compared with $\omega$ since the
shear falls off more rapidly than the rotation as the universe
expands (\cite{haw69,ell73}) and the isotropy of the CMB 
restricts the shear more
strongly. The conservation of energy and angular momentum gives (\cite{ell73})
\begin{eqnarray}
    \dot\rho=-(\rho+p)\Theta,~~~~\omega\rho a^5={\rm const}.
\end{eqnarray}
Especially, for dust we have $\rho_d\propto a^{-3}$ and $\omega_d\propto
a^{-2}$, for radiation we have  $\rho_r\propto a^{-4}$ and  $\omega_r\propto
a^{-1}$ (while in general $\sigma$ falls as $\sigma\propto a^{-3}$
(\cite{haw69})). Before the decoupling epoch $z_{dec}\sim 10^3$, the dust and the
radiation interact with each other strongly, they can be treated as
one unique fluid and have one unique angular velocity. After the
decoupling, the dust and the radiation evolve separately, they have
their own angular velocities which evolve according to different laws:
$\omega_d\propto a^{-2}$ and $\omega_r\propto a^{-1}$. Since today the
universe is dominated by dust, we take $\omega_d$ as the angular
velocity of the universe though the radiation may have a more large
angular velocity $\omega_r\sim z_{dec}\omega_d$. For the dust fluid we
have $A^a=0$ because the dust flows along geodesics. Neglecting the
shear term which is assumed to be sufficiently small, then the first
integration of Eq. (2) for dust gives
\begin{eqnarray}
    H^2\equiv\left({\dot a \over a}\right)^2={8\pi G\over 3}\rho -{2\over 3}\omega^2
    -{\kappa\over a^2},
\end{eqnarray}
where $\omega=\omega_d$, $\kappa$ is the integral constant which can
be made to be $+1$, $0$, or $-1$ by rescaling. It should be
remembered that though Eq. (4) describes the motion of the dust fluid,
exactly $\rho$ is the total mass density of dust and radiation since
the right hand side of Eq. (2) comes from $-R_{ab}u^au^b$ and the
Einstein equation (\cite{ciu95}). $\omega$ and $\rho$ can be written as
$\omega=\omega_0(1+z)^2$, $\rho=\rho_{d0}(1+z)^3+\rho_{r0}(1+z)^4$.
Because the Einstein-de Sitter model has provided sufficiently good
description of the universe since decoupling, we expect that
$\omega_0^2$ should be sufficiently small compared with $G\rho_{d0}$.

Now consider the formation of galaxies in a rotating and expanding universe. 
At some early epoch there is some density fluctuation in a region, then the 
expansion of the matter inside and around the region begins to be increasingly 
decelerated. Eventually the matter may stop expanding and begin to collapse 
and 
form a galaxy. For simplicity, we assume that the fluctuation is spherically 
symmetric. Then the region containing the proto-galaxy or the matter destined 
to form a galaxy should also be spherically symmetric. Suppose the original
mass and angular momentum of the proto-galaxy  are $M$ and $J_i$, 
the original radius 
of the proto-galaxy is $r_i$, then $J_i=2Mr_i^2\omega_i/5$, where
$\omega_i$ is the 
angular velocity of the universe at that epoch. This angular momentum is 
relative to gyroscopic frames. Another kind of useful local frames are 
galactic frames which our usual measurements are made relative to, 
by definition which co-rotate with the global rotation  
and whose origins are fixed at galactic centers. Certainly the original 
angular momentum relative to galactic frames is zero. After the galaxy 
has formed, the galaxy rotates relative to the galactic frames, which is
caused by the Coriolis force or the conservation of angular momentum, 
just like the formation of cyclones on the ground of the earth. At any 
epoch after the galaxy has formed, the angular momentum of the galaxy 
relative to the 
gyroscopic frames is $J_f=J+\beta Mr_f^2\omega_f$, where $J$ is the angular
momentum of the galaxy relative to the galactic frames, $r_f$ is the 
radius of the galaxy , $\omega_f$ is the angular velocity of the universe, 
and $\beta$ is a parameter determined by the distribution of the mass of
the galaxy. Using $\omega\propto(1+z)^2$, $\rho_d\propto(1+z)^3$, 
and $M = 4\pi\rho_{di}r_i^3/3$ (it is usually assumed that galaxy
formation takes place after the decoupling), the conservation of 
angular momentum $J_i=J_f$ leads to
\begin{eqnarray}
    J={2\over5}\left({3\over4\pi\rho_{d0}}\right)^{2/3}\omega_0M^{5/3}-\beta
    r_f^2(1+z_f)^2 \omega_0M.
\end{eqnarray}
For $z_f$ not too larger than 1, the second term
in the right hand side of Eq. (5) is usually sufficiently small 
compared with the first term, then we have
\begin{eqnarray}
    J\simeq kM^{5/3}, ~~~k={2\over5}\left({3\over4\pi\rho_{d0}}
    \right)^{2/3}\omega_0,
\end{eqnarray}
which is consistent with the empirical relation in Eq. (1). Thus the
global rotation of the universe gives a natural explanation of the
observed correlation between the angular momentum and mass of galaxies.

By studying the correlation between the angular momentum and 
mass of galaxies, it should be able to find the angular velocity of
the universe. The correlation for spiral galaxies has been investigated in detail
(\cite{nor73,dai78,car82,abr85}). 
It seems a suitable value for $k$ is $\sim0.4$ (in CGS units). 
Taking $\rho_{d0}=1.88\times10^{-29}\Omega h^2$ g cm$^{-3}$ ($\Omega$
is the density parameter of dust and $h$ is the Hubble constant in units
of $100$ km s$^{-1}$ Mpc$^{-1}$) and choosing $h = 0.75$ and 
$\Omega = 0.01$ (This value is measured
dynamically for the rich clusters of galaxies (\cite{pee93})), we have
\begin{eqnarray}
    \omega_0\simeq6\times10^{-21}{\rm rad~s^{-1}}\simeq2\times10^{-13}
    {\rm rad~yr^{-1}},
\end{eqnarray}
which is consistent with the value obtained by Birch (1982) and just
within the CMB limits for the Bianchi IX models (\cite{ciu95}). The result is
also consistent with that of Obukhov (1992) and Korotky and Obukhov (1994)
if we interpreted the angular velocity they obtained as the
angular velocity of the radiation according that in their models the angular
velocity decays as $\omega\sim a^{-1}$.

Let us turn to discuss the relation between the 
primordial density fluctuation and the formation of galaxies in a rotating 
and expanding universe. Consider a spherical shell with initial radius $r_i$ 
containing the spherically symmetric primordial density fluctuation with  
contrast $\delta_i=\delta\rho_i/\rho_i$ ($0<\delta_i\ll 1$), where
$\rho_i$ is the initial average mass density
of the universe and $\delta\rho_i$ is the fluctuation. 
When the density fluctuation 
appears, the shell and its interior (and the part of its exterior near the
shell) decrease the speed of expansion and are gradually separated from the
other parts of the universe to form an isolated system. The mass contained
in the shell is $M=4\pi\rho_ir_i^3(1+\delta_i)/3$, which is supposed to be 
constant during the evolution (it is a good approximation if the shell is not
very near the center of the fluctuation). Consider a mass element on the 
equator of the shell. When the system becomes isolated, the motion of the
mass element is equivalent to that of a particle with unit mass in the potential
$U^{(e)}(r)=-GM/r+\omega_i^2r_i^4/(2r^2)$ with the initial conditions
$r=r_i$ and $\dot r=H_ir_i$ at $t=t_i$, where $t_i$ is the time when
the fluctuation takes place. The total conserved energy of the
particle is
$\varepsilon^{(e)}\simeq-\vartheta\delta_iH_i^2r_i^2/(2(1+\vartheta))$
where Eq. (4) has been used and the $\kappa$ term has been dropped as
usual. The parameter $\vartheta$ is defined by
\begin{eqnarray}
    \vartheta\equiv{3\delta_i\over ({\omega_0/
    H_0})^2(1+z_i)}-1,~~~\vartheta>-1,
\end{eqnarray}
which describes the strength of the density fluctuation that takes place
at the redshift $z_i$. The solution is bound and there exists the turn-around
point where the mass element stops expanding and begins to collapse,
if $\varepsilon^{(e)}$ is negative or $\vartheta>0$. Under this
condition, the solution is
\begin{eqnarray}
    {r\over r_i}\simeq{1\over2}{1+\vartheta\over
    \vartheta\delta_i}(1-e\cos\xi), ~~~{t\over t_i} \simeq
    {3\over4}\left({1+\vartheta\over \vartheta\delta_i}\right)^{3/2} (\xi-e\sin\xi),
\end{eqnarray}
where $H_i\simeq2/(3t_i)$, $\delta_i\ll1$, and $\omega_i^2/H_i^2\ll1$
(For $z_i\sim10^3$ we have $\omega_i^2/H_i^2\sim10^{-3}$) have been
used, and $e=[1-12\vartheta\delta_i^2/(1+\vartheta)^2]^{1/2}$. The
collapse time in the equatorial direction is
\begin{eqnarray}
t_c^{(e)}\equiv t(\xi=2\pi)-t_i\simeq\left[{3\pi\over2\delta_i^{3/2}}
\left({1+\vartheta\over \vartheta}\right)^{3/2}-1\right]t_i.
\end{eqnarray}
For a mass element in the polar direction (the direction of the
rotation), when the 
system becomes isolated, the motion is equivalent to that of a particle with
unit mass in the potential $U^{(p)}(r)=-GM/r$ with the initial 
conditions $r=r_i$ and $\dot r=H_ir_i$ at $t=t_i$. The total
energy is $\varepsilon^{(p)}\simeq-(3+\vartheta)\delta_iH_i^2r_i^2/
(2(1+\vartheta))$ which is always negative. The solution 
is always bound and the turn-around point always exists. The solution is
\begin{eqnarray}
    {r\over r_i}\simeq{1\over2}{1+\vartheta\over(3+
    \vartheta)\delta_i}(1-\cos\eta), ~~~{t\over t_i} \simeq
    {3\over4}\left[{1+\vartheta\over(3+\vartheta)\delta_i}\right]^{3/2} (\eta-\sin\eta).
\end{eqnarray}
The collapse time in the polar direction is
\begin{eqnarray}
    t_c^{(p)}\equiv t(\eta=2\pi)-t_i\simeq\left[{3\pi\over2\delta_i^{3/2}}
    \left({1+\vartheta\over 3+\vartheta}\right)^{3/2}-1\right]t_i.
\end{eqnarray}
We find that $t_c^{(e)}\gg t_c^{(p)}$ if $0<\vartheta\ll1$.

There are three possible evolution 
results depending on the parameter $\vartheta$: 
\begin{itemize}
  \item $\vartheta>1$ or
    $\vartheta\sim1$.  Then $t_c^{(e)}\sim t_c^{(p)}$, the matter in the 
    equatorial and polar directions collapses and reaches
    dynamical equilibrium almost simultaneously. The objects so formed 
    are in complete equilibrium in both the equatorial and the polar 
    directions, which should have compact shapes. Such objects are just 
    like elliptical galaxies, the formation of 
    elliptical galaxies may therefore belong to such a case.
  \item $0<\vartheta\ll1$. Then $t_c^{(e)}\gg t_c^{(p)}$, the matter in the equatorial 
    direction collapses and reaches dynamical equilibrium sufficiently later than
    that in the polar direction. When the matter in the polar direction has stopped
    collapsing and has reached the equilibrium, the matter in the equatorial
    direction is still flowing into the core and is rotating around the core.
    The matter in the polar direction is in complete equilibrium, while the matter 
    in the equatorial direction is in quasi-equilibrium. The objects so formed are
    not as compact as that in the first case and are just like spiral galaxies, 
    the formation of spiral galaxies may therefore belong to such a case.
\end{itemize}
These two cases provide a
natural mechanism accounting for the formation of spiral galaxies and elliptical
galaxies. A direct corollary is that the distribution of the average mass density
of spiral galaxies should concentrate within a narrow range, the average mass
density of elliptical galaxies should scatter in a more wide range and should
be more large. This is consistent with the observations.
\begin{itemize}
  \item $-1<\vartheta\leq0$. In such a case there is no bound
    solution in the equatorial direction, the matter in this direction will expand
    forever though the expanding speed decreases with time, even when the
    matter in the polar direction has collapsed and has reached equilibrium. As the results,
    only two dimensional bound structures can be formed, which can be regarded as 
    proto-walls and provide natural seeds for the formation of wall structures in 
    the universe. The surrounding matter and galaxies are drawn towards a proto-wall
    to form a wall structure. The scale $L$ of the wall is approximately equal to the
    diameter $D^{(w)}$ of the proto-wall, which can be estimated by 
    $L\sim D^{(w)}\sim D_i^{(w)}(1+z_i)$, because the
    proto-wall can be approximately regarded as expanding with the universe in
    the equatorial direction. For a typical spiral galaxy, its original diameter
    can be estimated by $D_i^{(s)}\sim D_0^{(s)}(\omega^{(s)}/\omega_0)/(1+z_i)$
    due to the conservation of vorticity, where $D_0^{(s)}$ is the present diameter of the 
    spiral galaxy and $\omega^{(s)}$ is the angular velocity at $r\sim
    D_0^{(s)}/2$. Then $L\sim10D_0^{(s)}(\omega^{(s)}/\omega_0)^{1/2}\sim10^2$
    Mpc, if we take $D_i^{(w)}\sim10D_i^{(s)}$, $\omega^{(s)}\sim10^{-16}$ rad
    s$^{-1}$, $\omega_0\sim10^{-20}$ rad s$^{-1}$, and $D_0^{(s)}\sim10^2$
    kpc. This scale has the same 
    order as that of the Great Wall.
\end{itemize}

How galaxies get their angular momentum during their formation
is an interesting and challenging problem in cosmology. Some people have found
and discussed the similarity of spiral galaxies to turbulent eddies, and suggested 
that the primordial turbulence may lead to the formation of galaxies and may be
the origin of the rotation of galaxies (\cite{von51,gam52}).
But detail investigations have revealed that the primordial turbulence picture
should fail since the turbulence could not have been kept for a long time against
the dissipation (\cite{jon72,jon73}). Other people 
have suggested that galaxies
acquire their angular momentum as they form by the tidal torques of
neighboring proto-galaxies (\cite{hoy51,pee69} \& 1971; 
\cite{bar87}), but it seems difficult to
explain the empirical relation in Eq. (1) in this picture. In the
scenario of the global rotation, the Coriolis force 
in the galactic frames makes galaxies to rotate automatically when they form,
galaxies get their angular momentum from the global rotation of the universe
due to the conservation of angular momentum. Galaxies rotate because the universe 
rotates. Differing from the primordial turbulence, this rotation can be kept due to
the conservation of angular momentum and the dissipation cannot make the rotation
to stop. In such a scenario, the empirical relation in Eq. (1) can be
explained naturally. One may expect that in such
a scenario the spins of galaxies should not distribute in the sky randomly, there should
be a dipole anisotropy along the direction of the global rotation. As
mentioned in the beginning of this paper, such kind of anisotropy in
the distribution of the spin of galaxies has been found at different
levels. Here I point out that the derivation from spherical symmetry
of proto-galaxies before they collapse may weaken the alignment of
the spin of galaxies which makes it very difficult to observe the
correlation of the orientation of galaxies. One can 
imagine that a proto-galaxy may be highly asymmetric, the surface containing the matter 
destined to end up in a single galaxy may have a very irregular shape
(\cite{pee80}).
The moment of inertia tensor of such an object is usually very complex compared with 
that of a sphere, then in general the angular momentum of the proto-galaxy should not 
take the same direction as the angular velocity. When the proto-galaxy rotates and 
expands together with the universe, its angular velocity is equal to that of the
global rotation in both magnitude and direction. The rotation of the universe makes
the angular momentum of the proto-galaxy, which is not aligned with
the angular velocity 
with the fixed direction, to precess about the axis of the rotation. 
The magnitude of the
angular momentum is constant during the precession. When the proto-galaxy becomes 
separated from the global rotation and expansion of the universe, and begins to 
collapse to form a galaxy, the interaction with its surroundings should become more
and more weak and eventually negligible. Its angular momentum gradually becomes 
constant in both magnitude and direction. In general the direction of the angular
momentum is not aligned with the global rotation. It should be determined by the
shape of the proto-galaxy and the time when the proto-galaxy becomes an isolated
system. Its distribution in space can be expected to be almost random,
instead of a strong dipole distribution. As the galaxy evolves , 
the dissipation processes
inside it cause that the component of its angular velocity perpendicular to the 
angular momentum gradually vanishes, eventually the galaxy rotates
about the direction of its angular momentum. The influence of the
shape of the proto-galaxy on the
formulae in Eq. (6) can be estimated dimensionally: Let $l_i$ be some linear scale
of the proto-galaxy, then the mass $M\sim\rho_{di}l_i^3$, the moment of
inertia $I\sim Ml_i^2\sim M^{5/3}\rho_{di}^{-2/3}$, and the angular
momentum $J\sim I\omega_i\sim
M^{5/3}\rho_{d0}^{-2/3}\omega_0$. Therefore $J\simeq kM^{5/3}$  still
holds,  with $k\sim\rho_{d0}^{-2/3}\omega_0$. The initial shape 
of the proto-galaxy only affects the numerical factor in $k$. However,
these do not seem to strongly influence the estimation of the
order of magnitude of the angular velocity of the universe. And the
scattering of the observational data around Eq. (6) may just reflect
the effect of the original shape of proto-galaxies.

\acknowledgments
I have benefited a lot from the discussions with R. A. Matzner and
R. M. Wald on some issues, here I am very grateful to them.

\end{document}